\documentclass[12pt,preprint]{aastex}
\shorttitle{New Findings for the BP of the CMa}
\shortauthors{Powell et al.}
\begin{document}
\title{New Findings for the Blue Plume Stars in the Canis Major Over-Density}
\author{W. Lee Powell Jr. \altaffilmark{1,2}, Ronald Wilhelm \altaffilmark{2}, and Kenneth W. Carrell\altaffilmark{2,3}}
\affil{Department of Physics, Texas Tech University,
    Lubbock, TX 79409-1051}
\author{Amy Westfall}
\affil{McDonald Observatory, Fort Davis, TX}
\altaffiltext{1}{Visiting Astronomer, Cerro Tololo Inter-American Observatory.}
\altaffiltext{2}{Visiting Astronomer, McDonald Observatory.}
\altaffiltext{3}{Visiting Astronomer, Kitt Peak National Observatory.
KPNO and CTIO are operated by AURA, Inc.\ under contract to the National Science
Foundation.}
\begin{abstract}
We obtained new UBV photometry and spectroscopy of Blue Plume (BP) stars near the center of the Canis Major Over-Density (CMa).  We combined analyses of color-color diagrams with a new comparison of the hydrogen Balmer-line profile to the reddening-free Q parameter to improve the reddening and extinction estimates for this low-latitude, differentially reddened, area of the sky.  Results of our stellar parameter analysis for B/A spectral type stars associated with the BP show that the majority of the stars have main-sequence surface gravities placing them at an average heliocentric distance of $<$D$>$ = $6.0 \pm 2.7$ kpc.  This distance is more consistent with membership in the intervening Perseus spiral arm and strongly suggests that the BP stars are not associated with the other stellar populations previously reported to make up the CMa.  This result casts serious doubt on the proposed dwarf galaxy origin for the CMa.
\end{abstract}
\keywords{Galaxy: structure --- galaxies: interactions --- Galaxy: stellar content --- galaxies: individual (Canis Major)}
\section{Introduction}
The nature of the CMa, discovered by \citet{mart2004}, remains a controversial subject.  Whether the CMa is a distinct, disrupting dwarf galaxy or a feature of the Galactic disk remains open to debate.  Previous studies using either 2MASS or broadband Johnson-Cousins photometry have shown indications that over-densities in the red giant branch (RGB), red clump (RC) and old main-sequence (OMS)(Age = 4-10 Gyrs) exist, centered near l = 240, b = -7.  Each of these populations lie red-ward of the thin/thick disk turnoff and are identified using statistical methods which compare expected model densities to observations.  This type of analysis is necessitated because of the enormous number of foreground/background stars along the line-of-sight for these samples.  Results from these investigations are confounded by the difficulty in constraining the differential reddening which is prevalent in this portion of the sky.  Further, the use of smooth stellar models which may not accurately account for the Galactic warp and flare allow serious consideration that the CMa is simply a natural feature of the disk of the Galaxy \citep{Moma2006}.

It is only the young main-sequence (YMS) of the BP that stands out as being virtually free of field contamination, and it is the one CMa component which can be more cleanly modeled in order to constrain the distance to the CMa. The BP present in the central CMa fields have been studied photometrically by \citet{butl07}.  \citet{dejo2007} successfully fit the BP sequence using theoretical isochrones and predicted a distance to the BP of 9-10 kpc (depending on the assumed [Fe/H]) and suggested that it was consistent with the distance of both the OMS and the CMa.  This result, however, depends on the correct determination of reddening, and the location of the MS turn-off for the BP population.  Without the strength of argument supplied by the BP stars in the YMS, the case for the galactic origin of the CMa is severely diminished.

In this letter we present new results from our spectroscopic study of BP stars along the line-of-sight to the CMa.  A full paper will follow that provides complete details of our methods and results, including new spectra from a pending observing run at CTIO on the 4m Blanco telescope.

\section{Observations}

\subsection{Photometry}

Our study began with initial photometry obtained at McDonald Observatory (McD) using the 0.8m telescope.  We used this photometry to choose candidate BP stars for preliminary, low-resolution, spectroscopy using the Large Cassegrain Spectrograph (LCS) on the 2.7m at McD.  These data helped us understand the populations present in the CMa.  Based on these data, we applied for time at CTIO through NOAO.  The photometry data in this paper were then obtained using the 0.9m telescope at CTIO 24-29 November 2006.  UBV images were obtained for sixteen 13.5' fields under photometric conditions.  These data were reduced using standard techniques, including flat field and bias corrections, using quadproc in IRAF.  \citet{land92} standard fields were observed to calibrate the photometry.  Typical residuals in the solution were 1\% in V and B, and a few percent in U.  Color-magnitude diagrams (CMDs) and color-color (c-c) plots were made from the photometry data.  These were used to select BP stars as targets for future spectroscopy, and to estimate the reddening.  A representative CMD and c-c plot are shown in figure 1.  The targets for spectroscopy were drawn from a 3x3 grid of the photometry fields, or a field 40' on a side, centered on l = 238.65, b = -7.52.
\begin{figure}[!ht]
\epsscale{0.75}
\plotone{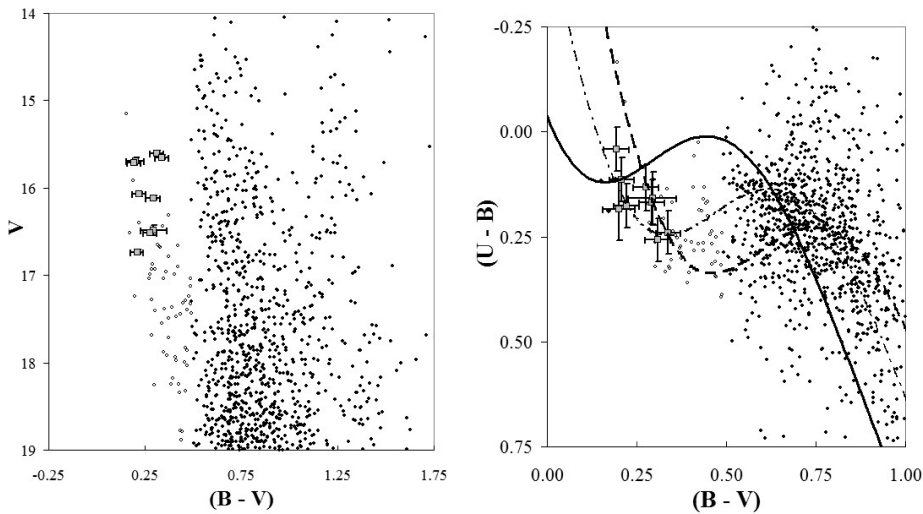}
\caption{A non-reddening corrected CMD and c-c plot for a representative field.  The stars with spectra are shown
 with error bars.  On the c-c plot, the ZAMS on the left is not reddened, the middle is reddened to 0.175
 and the ZAMS at right to 0.3.  Some stars are fit by more than one of the curves due to the
 degeneracy in the ZAMS fitting method.\label{fig1}}
\end{figure}

\subsection{Spectroscopy}

We obtained spectra for 58 BP stars using the Hydra Multi-Object Spectrograph on the WIYN 3.5m telescope at KPNO.  The spectra have a wavelength coverage from 3500 to 5400 $\AA$ with a resolution of R $\sim 2000$ and typical S/N $\sim 20$ at H$\delta$ and S/N $\sim 40$ at H$\beta$
.  The data were reduced using DOHYDRA in IRAF and helio-centric radial velocities were determined using FXCOR with synthetic spectral templates.  Analysis of the stellar parameters of Teff, Logg, and [Fe/H] used a combination of the Balmer-line widths and dereddened (see below) UBV photometry as described in \citet{wil99}.

It was not possible to use the CaII K line to establish metallicity, due to the extensive amount of contamination from interstellar Ca so near the Galactic plane.  Instead, metallicity for the stars with Teff $<$ 9750 (K) was determined using synthetic template comparison to metal-line regions in the observed spectra (Wilhelm et al. 1999).  The average metal abundance for these stars was found to be $<$[Fe/H]$>$ = -0.37 with a 1$\sigma$ spread of $\pm 0.18$.  The hotter stars in the sample were assigned [Fe/H] = -0.5 using comparison of theoretical isochrones from \citet{gia02} in
\begin{figure}[!ht]
\epsscale{0.7}
\plotone{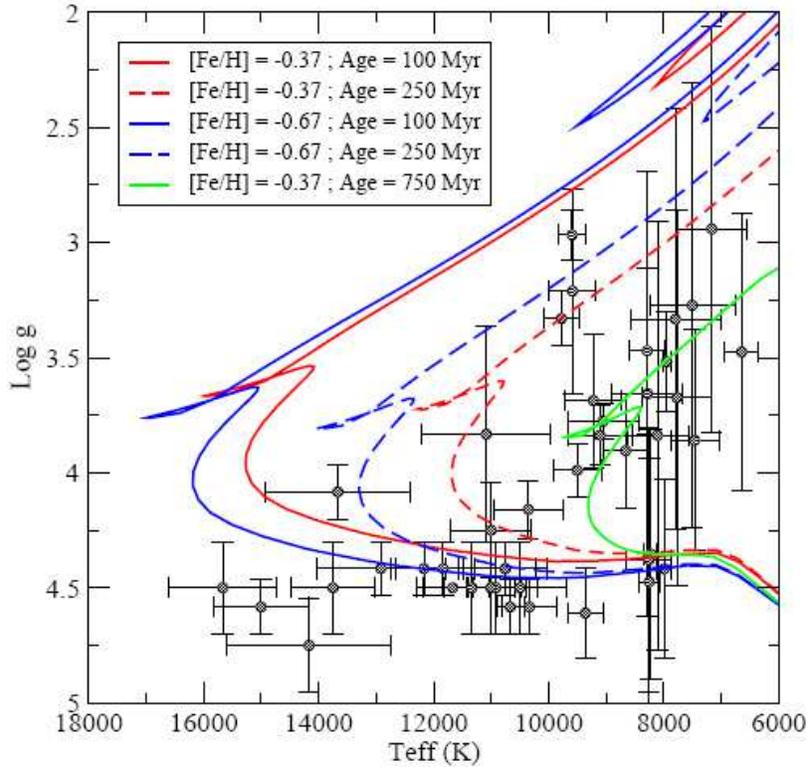}
\caption{The hotter stars in the sample have high surface gravities indicative of MS stars and not the MS turn-off as might be expected from the CMD.  The isochrone fits suggest an age spread of at least 650 Myrs.\label{fig2}}
\end{figure}
the Teff/Logg plane.  See figure 2 below.

Figure 2 is a plot of our results for 44 stars with data of sufficient quality for analysis.  We find that the majority of hot stars in this sample have Log g values consistent with main-sequence stars and not the lower surface-gravity main-sequence turn-off.  Distances were computed using absolute magnitudes, the theoretical isochrone fits to the Teff-Logg plane and the extinction corrected V magnitudes.

\section{Dealing with differential reddening}

The CMa has a low galactic latitude ($~$ -8 degrees) which results in heavy differential reddening.  Dealing with this reddening is crucial to understanding the populations present in the CMa.  The most common means of estimating the reddening is the use of the \citet{SFD1998} (SFD) dust maps, or the correction of these maps provided by \citet{bmb}. We instead used two methods to determine the reddening for individual stars.  Color-color plots were made for our fields using our photometry.  Since the BP is purported to be a YMS (De Jong et al.), we fit \citet{s-k1982} ZAMS curves to the c-c plots to account for the effect of differential reddening across the fields.  By shifting the ZAMS on a c-c plot until a given BP star fell on the curve, individual reddening could be determined.  A degeneracy arose since there can be two E(B - V) values that fit the ZAMS to a star, as can be seen in figure 1.  A second value for E(B - V) was obtained through analyzing the spectra independent of the ZAMS reddening value.  In this method to constrain the reddening we use a comparison between the H-beta line profile and the reddening-free Q-parameter given by:
\begin{displaymath}
Q  =  (U - B) - 0.72(B - V)
\end{displaymath}
The line profile was determined by measuring the width of the H-beta line at 13 positions starting with 10\% below the local-pseudo continuum and moving downward in increments of 3\% for 12 further measures.  The widths were analyzed using a $\chi^2$ goodness-of-fit of all 13 widths compared to identically produced profiles of synthetically generated H-beta lines created from \citet{kur93} Atlas9 model atmospheres and the spectral synthesis routine SPECTRUM \citep{gra94}.  Local minima where identified with a typical degeneracy found for cooler and hotter stars which lie on the either side of the A0 maximum.  Typical differences in the observed Q-value between the degenerate minima were on the order of 0.2 - 0.3.  This was sufficient to break the degeneracy and assign a true global minimum.  The observed Q-values were then compared to theoretical Q-values supplied by Atlas9 generated UBV colors and synthetic (B-V) colors were directly compared to the unreddening corrected, observed (B-V) in order to establish the E(B-V) for the star.  For the 48 stars with sufficient S/N in their spectra to apply both methods, the standard deviation of the differences in the E(B - V) estimates was 0.022.

We then compared our adopted E(B - V) with the values predicted by \citet{bmb} (BMB).  E(B - V) ranged from 0.02 to 0.555 for our method, and from 0.20 to 0.32 for the BMB method.  We found an average E(B - V) of 0.286 using our method and an average of 0.268 using the BMB method, showing that the two approaches give the same reddening estimate to within in the error estimate of 16\% claimed by SFD for their maps.  The standard deviation in E(B - V) for our method is 0.133, compared to 0.038 for the BMB method.  Clearly our method reproduces the expected average value, but is better able to account for the rapid changes in the value that result in this highly differential field.  The scale resolution of the SFD maps is 6'.  This shows that the maps are not able to adequately deal with these differential fields.

\section{Results and Conclusions}

Figure 3 shows a frequency histogram of the number of stars versus heliocentric distance and radial velocity.  Both of these plots have a result different than what has been attributed to the CMa.  First, the radial velocity average for our sample is 90 km/s rather than the 109 km/s found by Martin et al. for the RC.  Further, the BP is clearly not located at a distance of 8 or 10 kpc, but is rather distributed throughout the entire line of sight to the CMa.  The mean distance for our BP sample was around 6 kpc, with a large dispersion of 2.7 kpc.
\begin{figure}[!ht]
\epsscale{1.1}
\plottwo{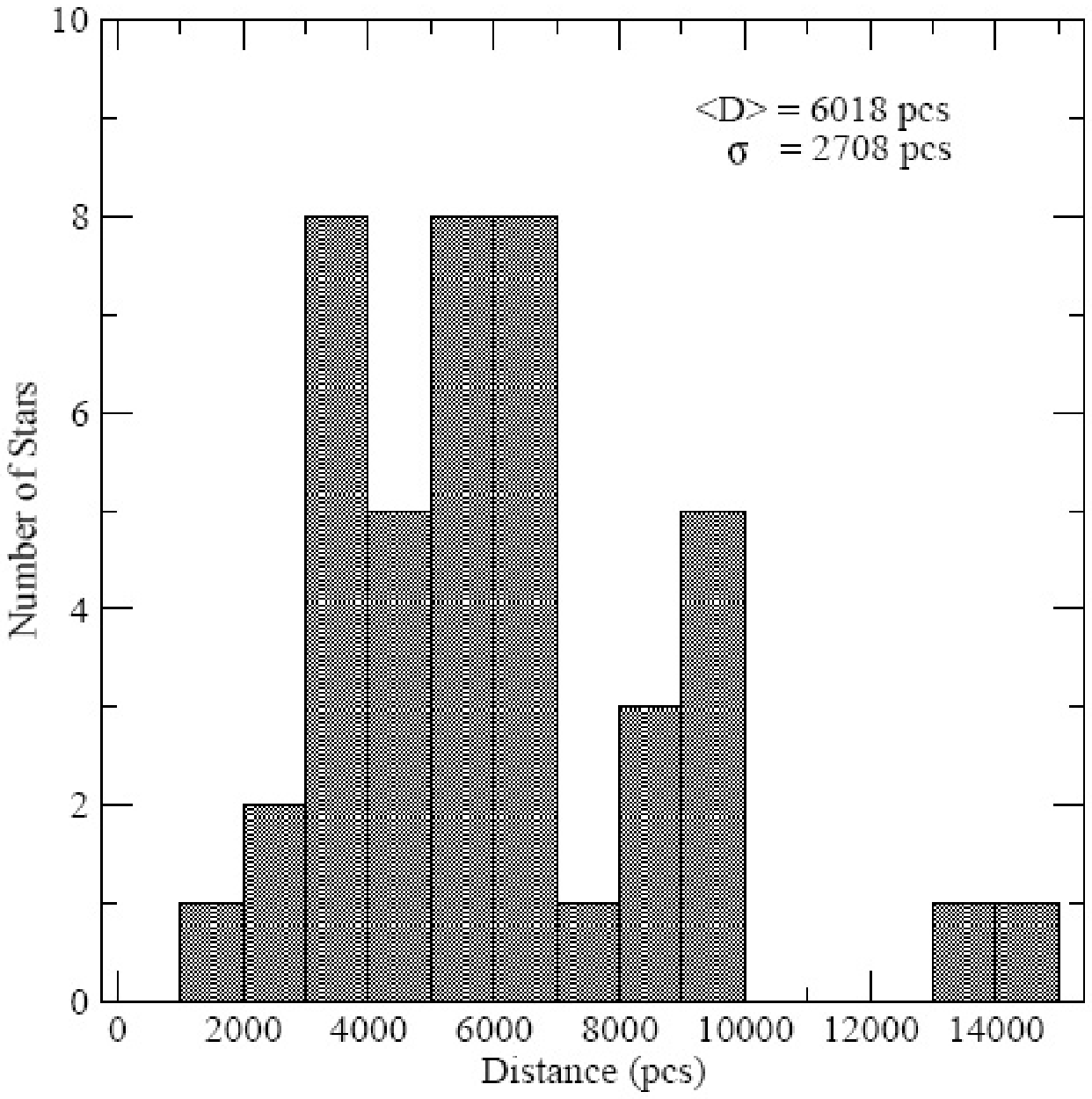}{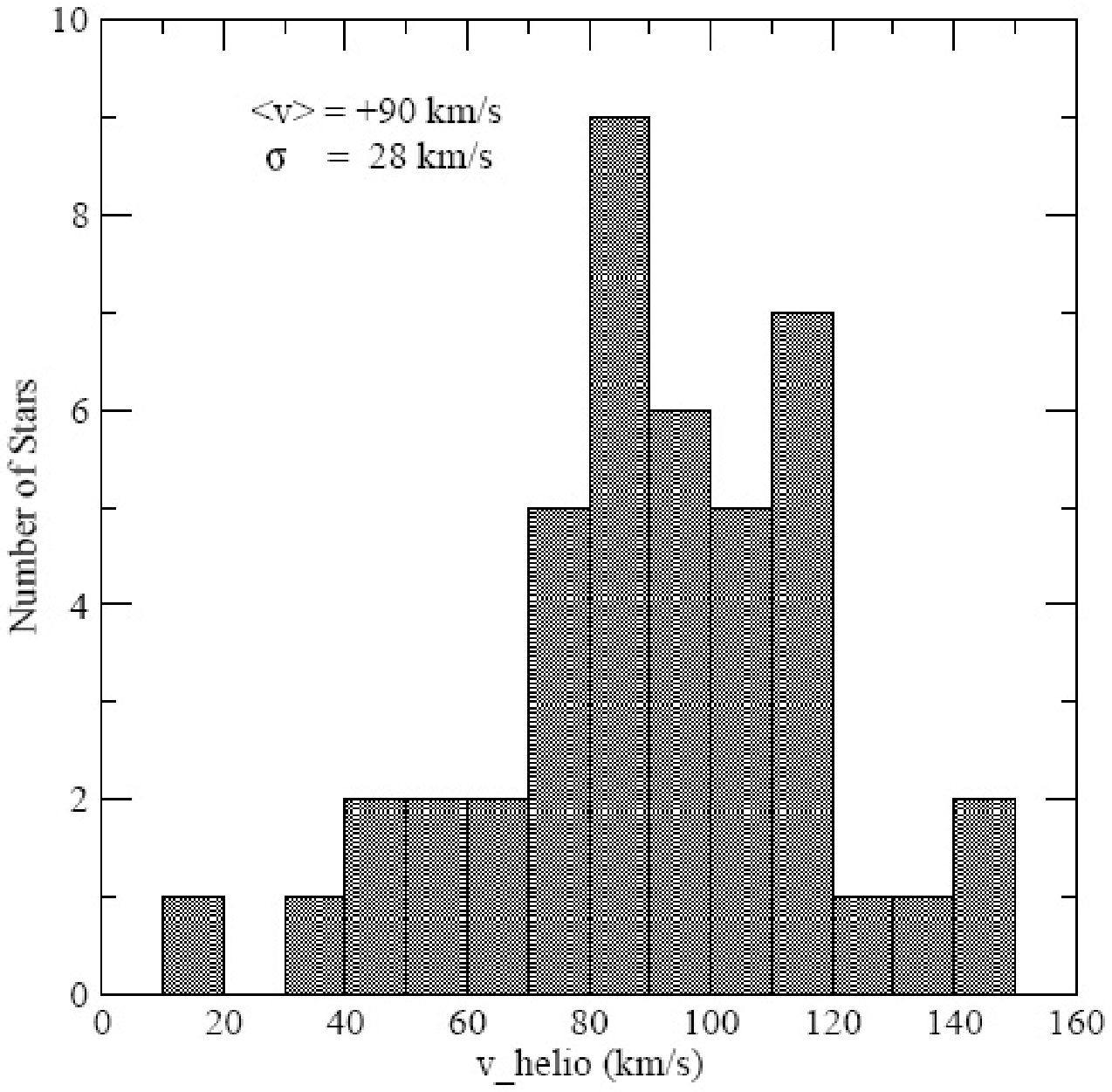}
\caption{The number of stars vs. distance and heliocentric radial velocity.\label{fig3}}
\end{figure}
This is at odds with the result found by De Jong et al., who predict a BP clump either at the distance of the CMa as predicted by Martin et al. or farther away and presumably associated with spiral arm structure in the Milky Way.  This is not surprising since the fitting techniques employed in their paper presume that the top of the BP coincides with the turn-off for the YMS.  As can be seen in Figure 2, the stars at the top of our BP are clearly MS dwarfs rather than turn-off stars.  The stars are also scattered at various distances making their analysis an inappropriate treatment of their data.  Finally, our result is in keeping with that of \citet{car05} who find similar BP stars at various Galactic longitudes which are very different from the line-of-sight longitude to CMa.

Our result also shows that the proper motion study of \citet{din05} is flawed by this same assumption.  They base their measurements on the assumption that the BP behaves as Martin et al. predict, namely that the clump lies at a distance of 8.1 $\pm 0.4$ kpc, and that it has a radial velocity average of 109 $\pm 4$ km/s.  We observe a BP with a dramatically different morphology and kinematic signature.

\citet{Moma2006} discuss the evidence that the CMa is simply due to the warp of the Galaxy, and refutes the arguments in favor of a dwarf galaxy origin for the CMa.  Figure 3 in this article shows a schematic view of the Milky Way based on various models.  The morphology we observe for the BP seems to fit well with these models if we presume that we are looking along the spiral arms.  Figure 16 in their paper shows a cut in the YZ plane of a warped and flared disk.  Our line of sight would trace a projected scale height of around one for much of the distance to the CMa.  In light of these arguments, we see no other viable explanation for our observed BP morphology than that we are seeing the stars of a warped Galactic disk.

{\it Facilities:} \facility{McD:0.8m ()}, \facility{Smith ()}, \facility{WIYN ()}, \facility{CTIO:0.9m ()}

\clearpage

\end{document}